# New approach to Fully Nonlinear Adiabatic TWM Theory

*Shunrong Qian*

I'm presenting a new elegant formulation of the theory of fully nonlinear adiabatic TWM (FNA-TWM) in terms of elliptic function here. Note that the linear case of SFG and DFG in the undepleted pump approximation described by the FVH representation has been exploited several years ago. For the sake of completeness, I present the pseudo-FVH representation to describe OPA. Moreover, I'm trying to display an overview of TWM processes and show that both the linear cases, the linear adiabatic SFG(DFG) and the linear OPA, are only the special cases of my theory. Finally I also point out that the geometric image of the so-called adiabatic basis acts as the geodesic line of the generalized Bloch sphere.

## I. Introduction

As one type of the simplest nonlinear optical process, three-wave-mixing (TWM) processes has been studied intensely and widely used in frequency conversion. In the slow-varying amplitude approximation and plane-wave approximation, the coupling equation in the continuous plane wave case writes:

$$\begin{cases} \dfrac{dE_j}{dz} = -i\Omega_j E_{3-j}^* E_3 \exp(-i\int_0^z \Delta k\ dz), j=1,2 \\ \dfrac{dE_3}{dz} = -i\Omega_3 E_1 E_2 \exp(i\int_0^z \Delta k\ dz) \end{cases} \quad (1)$$

where the coupling coefficient $\Omega_j = \dfrac{\omega_j}{n_j c}\chi_{eff}^{(2)}$ $(j=1,2,3)$. Note that if we set $E_1 = E_2 = \dfrac{\sqrt{2}}{2}E_\omega, E_3 = E_{2\omega}, \Omega_1 = \Omega_2 = \Omega_\omega$ and $\Omega_3 = \Omega_{2\omega}$, SHG can be described by equation (1).

Now set

$$E_j = \sqrt{\Omega_j}\ A_j \exp(-i\int_0^z \Delta k\ dz), j=1,2,3 \quad (2)$$

where $A_j\ (j=1,2,3)$ is the normalized amplitude. Subscribing (2) into (1) yields:

$$\begin{cases} \dfrac{d A_j}{d\xi} = i\,\Delta\Gamma\,A_j - i\,s\,A_{3-j}^*\,A_3, j=1,2 \\ \dfrac{d A_3}{d\xi} = i\,\Delta\Gamma\,A_3 - i\,s\,A_1 A_2 \end{cases} \quad (3)$$

where $s$ is the sign of effective nonlinear coefficient, and $\Omega_0 = \sqrt{\left|\prod_{i=1}^{3} \Omega_i\right|}$, $\xi = \Omega_0 z$, $\Delta\Gamma = \Delta k / \Omega_0$.

For simplicity, set $I_j = |A_j|^2$ and $\varphi_j = \arg(A_j)$, $j = 1, 2, 3$. then it's not hard to obtain the Manley-Rowe relations:

$$\begin{cases} K_j = I_j + I_3, j = 1, 2 \\ K_3 = I_1 - I_2 \end{cases} \quad (4)$$

where $K_j$ ($j = 1, 2, 3$) are Manley-Rowe constants.

In the special case when coupling coefficient $\Omega_j$ and phase-mismatch $\Delta k$ are both constant, the analytic solution has been presented by J. A. Armstrong et al. in 1962[1]. However, since the phase mismatch acts as a ghost to interfere the frequency conversion, several different methods were put forward to swept the ghost out, among which the theory of QPM is the most hopeful and powerful one to provoke researchers to spatial modulated the coupling coefficient $\Omega_j$ and phase-mismatch $\Delta k$ in different ways. An interesting way is the adiabatic method, firstly posted by H. Suchowski in 2008[2]. Since then, adiabatic method becomes more and more popular in nonlinear optics[3][4][5]. Moreover, the attempt to remove the undepleted pump approximation from TWM also hastened the theory of fully nonlinear adiabatic TWM posted by G. Porat and A. Arie in 2013[6]. Here the author want to present a new elegant formulation of that theory in terms of elliptic function without the introduction of Hamiltonian formulation. For completeness, to begin, the author decides to briefly present the SFG(DFG) and OPA in the undepleted pump approximation, which namely, in short, linear SFG(DFG) and OPA in this article.

**II. The Linear SFG(DFG) and OPA**
**A. The FVH representation**

According the theory of H. Suchowski, the Schrodinger equation of linear SFG writes:

$$i\frac{d}{d\xi}|\psi\rangle = \hat{H}_h |\psi\rangle \quad (5)$$

where $\hat{H}_h = \frac{1}{2}(\text{Re}(\kappa) \cdot \boldsymbol{\sigma}_1 + \text{Im}(\kappa) \cdot \boldsymbol{\sigma}_2 + \Delta k \cdot \boldsymbol{\sigma}_3)$, $|\psi\rangle = [A_i, A_s]^T$, $\kappa = -\frac{\chi_{eff}^{(2)}}{c}\sqrt{\frac{\omega_i \omega_s}{n_i n_s}} E_p$,

$A_i = \sqrt{\frac{n_i}{\omega_i}} E_i \exp(-\frac{1}{2}i\int_0^z \Delta k \, dz)$, $A_s = \sqrt{\frac{n_s}{\omega_s}} E_s \exp(\frac{1}{2}i\int_0^z \Delta k \, dz)$, and $\boldsymbol{\sigma}_1, \boldsymbol{\sigma}_2, \boldsymbol{\sigma}_3$ are

three Pauli matrices, writes:

$$\sigma_1 = \begin{pmatrix} 0 & 1 \\ 1 & 0 \end{pmatrix}, \sigma_2 = \begin{pmatrix} 0 & -i \\ i & 0 \end{pmatrix}, \sigma_3 = \begin{pmatrix} 1 & 0 \\ 0 & -1 \end{pmatrix}.$$

Clearly, the Schrodinger equation of linear DFG shares the same form with that of linear DFG.

Note that $|\psi\rangle$ can be described by Bloch vector $\vec{\rho} = \{U, V, W\}$ in the FVH representation, where

$$U = \text{Re}(A_i A_s^*), \quad V = \text{Im}(A_i A_s^*), \quad W = |A_i|^2 - |A_s|^2 \tag{6}$$

It's not hard to see that $\vec{\rho}$ indeed falls on the Bloch sphere whose evolution is described by the well-known Bloch Equation:

$$\frac{d\vec{\rho}}{dz} = \vec{\Omega} \times \vec{\rho} \tag{7}$$

where $\vec{\Omega} = \{\text{Re}(\kappa), \text{Im}(\kappa), \Delta k\}$ is the torque vector whose norm is equal to the instantaneous rotational angular velocity of $\vec{\rho}$ about $\vec{\Omega}$, i.e. $\sqrt{|\kappa|^2 + \Delta k^2}$.

**Table 1 The eigensystem of Hamiltonian** $\hat{\mathbf{H}}_h$

| Eigenvalues | $\lambda_1 = -\frac{1}{2}\sqrt{(\Delta k)^2 + \kappa^2}$ | $\lambda_1 = \frac{1}{2}\sqrt{(\Delta k)^2 + \kappa^2}$ |
|---|---|---|
| Eigenvectors | $|\psi_1\rangle = \{\sin(\theta), -\cos(\theta)\}$ | $|\psi_2\rangle = \{\cos(\theta), -\sin(\theta)\}$ |

Assume that the coupling coefficient $\kappa$ is real. The eigenvector $|\psi_2\rangle$ shown in **Table 1** is used for the adiabatic basis in the case of linear SFG, where $\theta$ is the mixing angle and $\delta = \tan(2\theta) = -\frac{\kappa}{\Delta k}$, while $|\psi_1\rangle$ in the case of the linear DFG.

Clearly, to achieve the full frequency conversion from signal to idle wave, the mixing angle $\theta$ needs to change from $\theta = 0$ to $\theta = \frac{\pi}{2}$ during the whole TWM process.

According to quantum adiabatic theorem, the mixing angle $\theta$ should satisfy the adiabaticity condition of linear SFG(DFG), which writes:

$$|\dot{\theta}| \ll \sqrt{\Delta k^2 + \kappa^2} \tag{8}$$

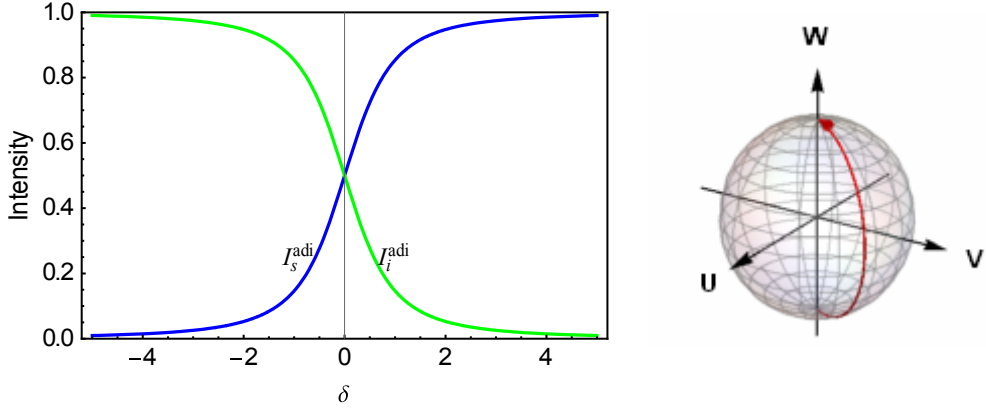

**Figure 1**  Adiabatic trajectory ($I_s^{adi}, I_i^{adi}$) of SFG and its image in Bloch sphere

## B. The Pseudo-FVH representation

In a similar way, the Schrodinger equation of linear OPA writes:

$$i\frac{d}{dz}|\psi\rangle = \hat{\mathbf{H}}_{nh}|\psi\rangle \tag{9}$$

where $\hat{\mathbf{H}}_{nh} = \frac{1}{2}(\text{Re}(\kappa)\cdot\tilde{\boldsymbol{\sigma}}_1 + \text{Im}(\kappa)\cdot\tilde{\boldsymbol{\sigma}}_2 + \Delta k\cdot\tilde{\boldsymbol{\sigma}}_3)$, $|\psi\rangle = [A_i, A_s]^T$, $\kappa = -\frac{\chi_{eff}^{(2)}}{c}\sqrt{\frac{\omega_i\omega_s}{n_i n_s}}E_p$,

$A_s = \sqrt{\frac{n_s}{\omega_s}}E_s\exp(\frac{1}{2}i\int_0^z \Delta k\, dz)$, $A_i = \sqrt{\frac{n_i}{\omega_i}}E_i\exp(\frac{1}{2}i\int_0^z \Delta k(\zeta)\, d\zeta)$, and $\tilde{\boldsymbol{\sigma}}_1, \tilde{\boldsymbol{\sigma}}_2, \tilde{\boldsymbol{\sigma}}_3$ are three pseudo-Pauli matrices, writes:

$$\tilde{\boldsymbol{\sigma}}_1 = \begin{pmatrix} 0 & -1 \\ 1 & 0 \end{pmatrix}, \tilde{\boldsymbol{\sigma}}_2 = \begin{pmatrix} 0 & i \\ i & 0 \end{pmatrix}, \tilde{\boldsymbol{\sigma}}_3 = \begin{pmatrix} 1 & 0 \\ 0 & -1 \end{pmatrix}$$

Note that $|\psi\rangle$ can be described by Bloch vector $\vec{\rho} = \{U, V, W\}$ in the FVH representation, writes:

$$U = \text{Re}(A_i A_s^*), V = \text{Im}(A_i A_s^*), W = |A_i|^2 + |A_s|^2 \tag{10}$$

Set $K = |A_s|^2 - |A_i|^2$. It's not hard to see that, $\vec{\rho}$ falls on one sheet of the biparted hyperboloid when $K > 0$, while on the cone when $K = 0$. We only focus on the former case, and call that surface pesudo-Bloch sphere.

Note that $\vec{\rho}$ also has its "Bloch Equation", writes:

$$\frac{d\vec{\rho}}{dz} = \vec{\Omega} \otimes \vec{\rho} \tag{11}$$

where $\vec{\Omega} = \{\text{Re}(\kappa), \text{Im}(\kappa), \Delta k\}$, the binary operator $\otimes$ represent a special product of two three-dimensional vectors, named pseudo-cross product, whose definition is:

$\vec{A} \otimes \vec{B} = \tilde{\varepsilon}_{i,j,k} A_j B_k \vec{\mathbf{e}}_i$, and $\tilde{\varepsilon}_{i,j,k}$ is a bit similar to the well-known Levi-Civita symbol $\varepsilon_{i,j,k}$, which writes:

$$\tilde{\varepsilon}_{i,j,k} = \begin{cases} 1 & ,(i,j,k)=(3,2,1),(1,3,2),(3,2,1) \\ -1 & ,(i,j,k)=(1,2,3),(2,1,3),(3,1,2) \\ 0 & ,otherwise \end{cases}$$

Such a "Bloch Equation", which can be derived from quantum Liouville equation using the algebraic properties of three pseudo-Pauli matrices introduced above, implies that the evolution of $\vec{\rho}$ is similar to Lorenz rotation in the special theory of relativity.

Suppose that $\kappa$ is purely imaginary with the positive imaginary part and $q = |\kappa|$. Then focus on the case when $|\Delta k| > q$. According to the theory of S. Ibáñez et al.[7], to completely describe the non-Hermitian Hamiltonian needs the biorthogonal eigen-equations:

$$\begin{cases} \hat{\mathbf{H}} |\psi_n\rangle = E_n |\psi_n\rangle \\ \hat{\mathbf{H}}^\dagger |\widehat{\psi}_n\rangle = E_n^* |\widehat{\psi}_n\rangle \end{cases} \quad (12)$$

where $|\widehat{\psi}_n\rangle$ and $|\psi_n\rangle$ are adjoint vectors, which satisfy the following orthonormal relations: $\langle \psi_m | \widehat{\psi}_n \rangle = \delta_{m,n}$.

**Table 2** The eigensystem of $\hat{\mathbf{H}}_{nh}$ and $\hat{\mathbf{H}}_{nh}^\dagger$ when $|\Delta k| > q$

| Eigenvalues | | $\lambda_1 = -\frac{1}{2}\sqrt{(\Delta k)^2 - q^2}$ | $\lambda_2 = +\frac{1}{2}\sqrt{(\Delta K)^2 - q^2}$ |
|---|---|---|---|
| Eigenvectors | $\hat{\mathbf{H}}_{nh}$ | $|\psi_1\rangle = \{i\sinh(\theta), \cosh(\theta)\}$ | $|\psi_2\rangle = \{\cosh(\theta), -i\sinh(\theta)\}$ |
| | $\hat{\mathbf{H}}_{nh}^\dagger$ | $|\widehat{\psi}_1\rangle = \{-i\sinh(\theta), \cosh(\theta)\}$ | $|\widehat{\psi}_2\rangle = \{\cosh(\theta), i\sinh(\theta)\}$ |

Note that the eigenvector $|\psi_2\rangle = \{\cosh(\theta), -i\sinh(\theta)\}$ shown in **Table 2** is certainty the adiabatic basis, where $\delta = \tanh(2\theta) = -\frac{q}{\Delta k}$. Represent the intensity of adiabatic signal and idle wave as a function of $\delta$:

$$I_s^{adi} = \frac{1}{2}\left(1 + \frac{1}{\sqrt{1-\delta^2}}\right), \quad I_i^{adi} = \frac{1}{2}\left(-1 + \frac{1}{\sqrt{1-\delta^2}}\right) \quad (13)$$

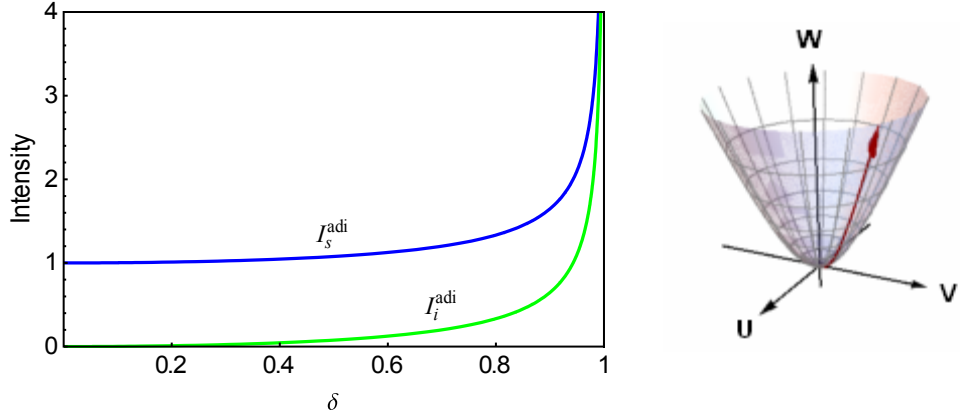

**Figure 2 Adiabatic trajectory ($I_s^{adi}, I_i^{adi}$) of DFG and its image in pseudo-Bloch sphere**

Clearly, to achieve the full frequency conversion from signal to idle wave, the mixing angle $\theta$ needs to change from $\theta = 0$ to $\theta = +\infty$ during the whole TWM process.

According to the non-Hermitian quantum adiabatic theorem, $\theta$ should satisfy the adiabaticity condition of linear SFG(DFG), which writes[7]:

$$\left|\dot{\theta}\right| << \sqrt{\Delta k^2 - \kappa^2} \tag{14}$$

Set the linear adiabatic parameter $r_l$ to be $r_l = \left|\dot{\theta}\right| / \sqrt{\Delta k^2 - \kappa^2} << 1$. One can see that as $\delta$ approaches 1, $r_l$ will become much larger than 1, and that's to say, the adiabatic trajectory would break down. Of course, a workaround may be super-adiabatic method[8][9], but there is no need to introduce it here. However, in the case of fully nonlinear adiabatic OPA, it would become better (see Section III.B).

Hitherto, two types of adiabatic processes have been introduced here, one is linear adiabatic SFG (DFG), and another is linear adiabatic OPA. In the next section, the reader may find that the two cases are both the special cases of fully nonlinear TWM process. One may also see that the geometric representation of the so-called adiabatic basis acts as the geodesic line of the generalized Bloch sphere.

### III. Fully Nonlinear Adiabatic TWM (FNA-TWM) Theory
### A. Fully Nonlinear Adiabatic SFG(including SHG) and DFG

Let's discuss adiabatic SFG first. Note that adiabatic SHG can be viewed as a special case of adiabatic SFG.

Assume that $K_1 \geq K_2 > 0$. According the Manley-Rowe relation (4), one can parameterize $A_j$ ($j=1,2,3$) by u:

$$A_1 = \sqrt{K_1}\,\text{dn}(u)e^{i\varphi_1}, \quad A_2 = \sqrt{K_2}\,\text{cn}(u)e^{i\varphi_2}, \quad A_3 = \sqrt{K_2}\,\text{sn}(u)e^{i\varphi_3} \quad (15)$$

where $\text{sn}(u), \text{cn}(u), \text{dn}(u)$ are Jacobi's elliptic functions sharing common parameter which writes $m = K_2/K_1$, and $u$ is a function of $\xi$. One should note that the least common period of the absolute value of elliptic functions introduced above is $T = 2\text{K}(m)$, where $\text{K}(m)$ represents the complete elliptic integral of first kind:

$$\text{K}(m) = \int_0^{\frac{\pi}{2}} \frac{d\phi}{\sqrt{1-m\sin^2\phi}}.$$

Define $J_\pm(u) = \dfrac{\text{sn}(u)\,\text{dn}(u)}{\text{cn}(u)} \pm \dfrac{\text{dn}(u)\,\text{cn}(u)}{\text{sn}(u)} \mp m\dfrac{\text{cn}(u)\,\text{sn}(u)}{\text{dn}(u)}$. Subscribing (15) into (3), and then separating the real and imaginary part, yields the coupling equation of $u$ and $\beta$:

$$\begin{cases} \dot{u} = s\sqrt{K_1}\sin(\beta) \\ \dot{\beta} = \Delta\Gamma - s\sqrt{K_1}\cos(\beta)J(u) \end{cases} \quad (16)$$

where $J(u) = J_-(u)$, $\beta = \varphi_1 + \varphi_2 - \varphi_3$. Since $J(u)$ is also a function with a period of $T$, i.e. $J(u+T) = J(u)$, we only need to focus on the following region: $R = \{(u,\beta) | 0 \le u \le T/2, -\pi \le \beta \le \pi\}$ (named the reduced $(u,\beta)$ plane).

Now let's calculate the stationary points: $(u^{(s)}, \beta^{(s)})$, which satisfying the systems of equations: $\dot{u} = \dot{\beta} = 0$. One can easily get the roots $\beta^{(s)}$ of the first equation of (16):

$$\beta^+ = 0,\ \beta^- = \pi \quad (17)$$

Having obtained the values of $\beta^{(s)}$, $u^{(s)}$ which says $u^+$ ($u^-$) corresponding to $\beta^+$ ($\beta^-$) respectively, can be obtained from solving the second equation of (16):

$$\Delta\Gamma - s\cos(\beta^{(s)})\sqrt{K_1}J(u) = 0 \quad (18)$$

For the sake of simplicity, set $s = 1$. Though the values of $u^+$ and $u^-$ can only be calculated numerically, one should note that according to equation (17) and (18), **Figure 3**(a) in fact shows us the adiabatic trajectory discussed below. Through spatial modulating $\Delta\Gamma$ from a small negative value to a large positive one increasingly, we can obtain the positive adiabatic trajectory $(u^+, \beta^+)$, which actually fits the process of SFG(SHG), since $I_3^{adi} \approx 0$ as $u^{(s)} \approx 0$ while $\Delta\Gamma \to -\infty$, and $I_3^{adi} \approx I_2^{(0)}$ as $u^{(s)} \approx T/2$ while $\Delta\Gamma \to +\infty$. Clearly, the parameter $u^{(s)}$ corresponds to mixing angles in linear case. To achieve the full frequency conversion from signal to idle wave, "the mixing angle" $u^{(s)}$ needs to change from $0$ to $T/2$ during the whole TWM process.

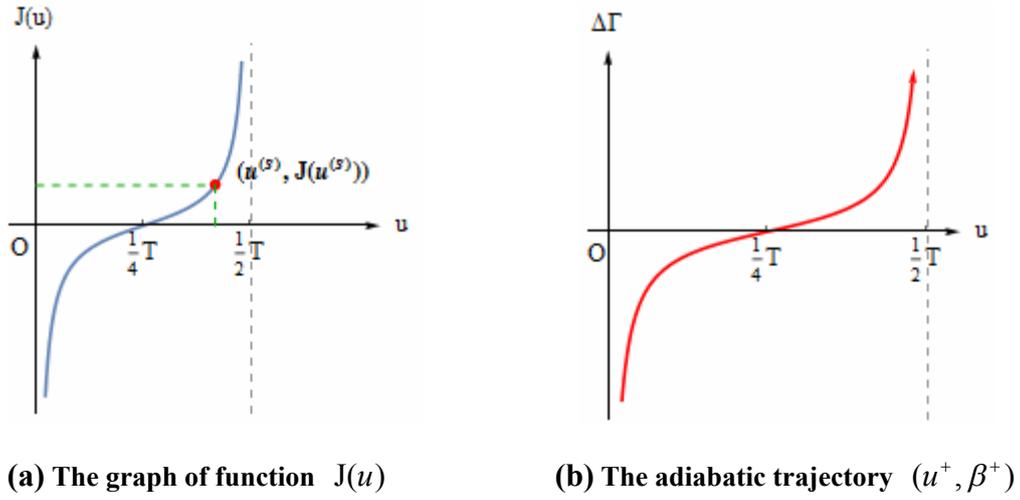

**(a)** The graph of function $J(u)$      **(b)** The adiabatic trajectory $(u^+, \beta^+)$

**Figure 3**

Now try to derive the adiabaticity condition. Firstly, suppose that there is a small displacement of $\Pi_0 = (u^{(s)}, \beta^{(s)})$, writes

$$\delta\Pi = (\delta u, \delta\beta) = (u, \beta) - (u^{(s)}, \beta^{(s)}) \tag{19}$$

In the first order approximation of $(\delta u, \delta\beta)$ at the point $(u^{(s)}, \beta^{(s)})$, (16) becomes:

$$\frac{d}{d\xi}\begin{pmatrix}\delta u \\ \delta\beta\end{pmatrix} = \begin{pmatrix} 0 & s\sqrt{K_1} \\ -s\sqrt{K_1}J'(u^{(s)}) & 0 \end{pmatrix}\begin{pmatrix}\delta u \\ \delta\beta\end{pmatrix} + \begin{pmatrix}\dfrac{du^{(s)}}{d\xi} \\ 0\end{pmatrix} \tag{20}$$

together with initial condition $(\delta u(0), \delta\beta(0)) = (0,0)$.

Following G. Porat and A. Arie, one can obtain that,

$$\delta u = \int_0^{\xi'} \cos(\Omega(\xi - \xi'))\frac{du^{(s)}}{d\xi}d\xi' \approx \frac{1}{\Omega}\sin(\Omega\xi)\frac{du^{(s)}}{d\xi} \tag{21}$$

where $\Omega = \sqrt{K_1 J'(u^{(s)})}$. Set the nonlinear adiabaticity condition to be $|\delta u| \ll 1$, then we obtain:

$$\left|\frac{du^{(s)}}{d\xi}\right| \ll |\Omega| \tag{22}$$

For the case when $m = 0$, the adiabaticity condition of SFG in terms of $u^{(s)}$ writes:

$$\left|\frac{du^{(s)}}{d\xi}\right| \ll \sqrt{\Delta\Gamma^2 + 4K_1} \tag{23}$$

Recall that in the undepleted pump approximation, the adiabaticity condition writes in a similar form of (26) (see Section II.A):

$$|d\theta/dz| \ll \sqrt{\Delta k^2 + \kappa^2} \tag{24}$$

Note that $\left|\dfrac{du^{(s)}}{d\Delta\Gamma}\right| = \left|\left(\dfrac{d\Delta\Gamma}{du^{(s)}}\right)^{-1}\right| = |\Omega|^{-2}$, one can obtain:

$$\left|\frac{d\Delta\Gamma}{d\xi}\right| \ll \frac{|\Omega|^3}{\sqrt{K_1}} \tag{25}$$

Set the nonlinear adiabaticity parameter to be $r_{nl}$:

$$r_{nl} = \sqrt{K_1} |\Omega|^{-3} \left|\frac{d\Delta\Gamma}{d\xi}\right| \ll 1 \tag{26}$$

The nonlinear adiabaticity parameter $r_{nl}$ characterizes the magnitude of the diabatic coupling of two different adiabatic passages, i.e. $(u^+, \beta^+)$ and $(u^-, \beta^-)$.

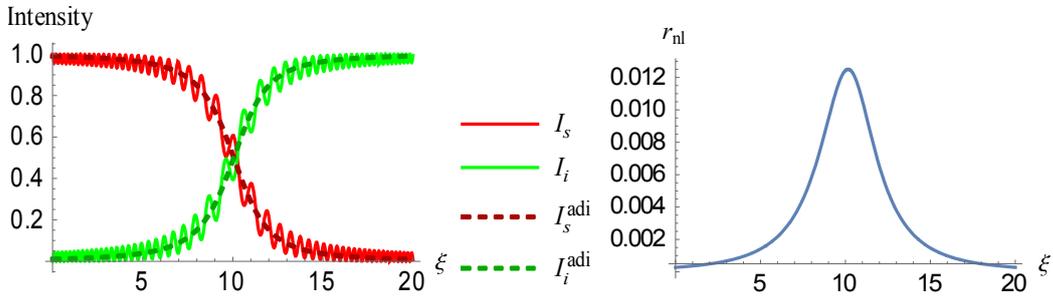

(a) The evolution of $I_s$ and $I_i$ in SFG    (b) Nonlinear adiabaticity parameter

**Figure 4 Adiabatic SFG with $K_1/K_2 = 10$ and $\Delta\Gamma = 3(\xi - 3)$**

Now that the general case as $m \in [0,1]$ have been shown, let's observe what on earth happens in the limiting cases i.e. when $m$ is at one of the end points of the interval $[0,1]$.

**(1)** When $m=0$, the pump wave never depletes, and the adiabatic trajectory $(u^+, \delta^+)$ writes:

$$I_1^{adi} = K_1, \quad I_2^{adi} = K_2 \cos^2(u), \quad I_3^{adi} = K_2 \sin^2(u) \tag{27}$$

which is the result shown in Section II.A.

**(2)** When $m=1$, as a case that can also describe SHG, the adiabatic trajectory $(u^+, \beta^+)$ writes:

$$I_\omega^{adi} = 2K \operatorname{sech}^2(u), \quad I_{2\omega}^{adi} = K \tanh^2(u) \tag{28}$$

where $K_1 = K_2 = K$.

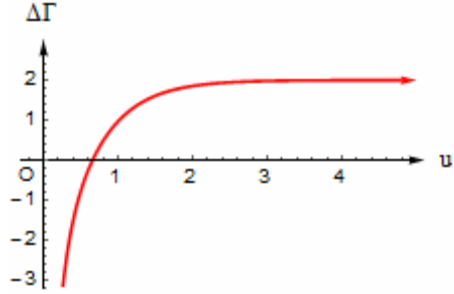

**Figure 5** The adiabatic trajectory $(u^+, \beta^+)$ of SHG

For the case when $u^{(s)} \approx 2$, on one hand, one can see from **Figure 5**(a) that, $\Delta\Gamma$ becomes steady at $2$, on the other hand, from **Figure 5**(b) that $du^{(s)}/d\Delta\Gamma$ becomes very large. Clearly, once $u$ is larger than 2, $r_{nl}$ becomes much larger than 1, and the adiabaticity condition (26) would be violated, so the adiabatic trajectory may break down, which is similar to the case of linear adiabatic OPA (see Section II.B).

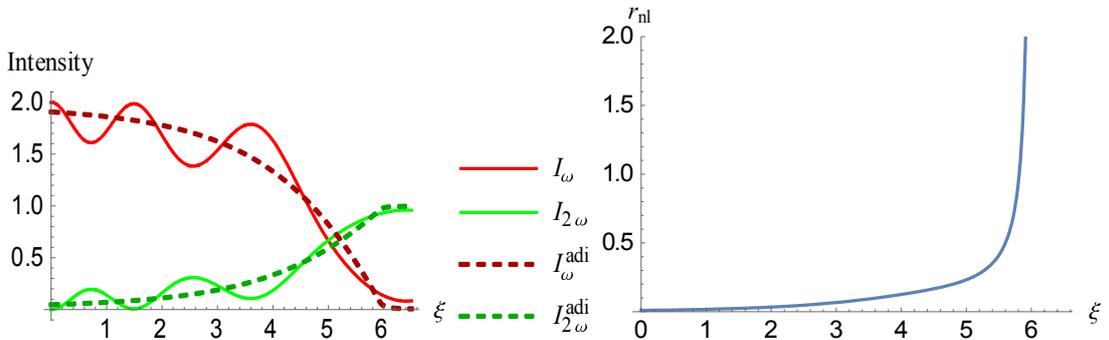

(a) The evolution of $I_s$ and $I_i$ in SHG    (b) Nonlinear adiabaticity parameter

**Figure 6 Adiabatic SHG with** $\Delta\Gamma = \xi - 4$

For the case of DFG, in a similar way, assume that $K_2 \geq K_1 > 0$. According the Manley-Rowe relation (4), we can parameterize $A_j\ (j=1,2,3)$ by u:

$$A_1 = \sqrt{K_2}\,\text{dn}(u)e^{i\varphi_1},\quad A_2 = \sqrt{K_1}\,\text{cn}(u)e^{i\varphi_2},\quad A_3 = \sqrt{K_1}\,\text{sn}(u)e^{i\varphi_3} \qquad (29)$$

where $\text{sn}(u), \text{cn}(u), \text{dn}(u)$ are Jacobi's elliptic functions whose parameter namely $m = K_1/K_2$, and $u$ is a function of $\xi$. Since the derivation of the adiabatic trajectory of DFG is completely similar to that shown above, the author chooses to omit it here.

**B. Fully Nonlinear Adiabatic OPA**

Following the approach showed above, let's turn to the theory of adiabatic OPA. Assume $K_2 \geq K_1 > 0$, and parameterize $A_j\ (j=1,2,3)$ by u:

$$A_1 = \sqrt{-K_3}\,\text{sn}(i u)e^{i\varphi_1},\quad A_2 = \sqrt{-K_3}\,\text{cn}(i u)e^{i\varphi_2},\quad A_3 = \sqrt{K_1}\,\text{dn}(i u)e^{i\varphi_3} \qquad (30)$$

where $\text{sn}(u), \text{cn}(u), \text{dn}(u)$ are Jacobi's elliptic functions sharing common parameter which writes $m = K_3/K_1$, and $u$ is a function of $\xi$.

Note that in the general case the least common period of elliptic functions introduced above is $T = 2(i\text{K}(m) + \text{K}'(m))$, where $\text{K}(m)$ is defined in Section III.A, and $\text{K}'(m) = \text{K}(1-m)$.

Subscribing (30) into (3), yields:

$$\begin{cases} \dot{u} = -s\sqrt{K_1}\cos(\beta) \\ \dot{\beta} = \Delta\Gamma - s\sqrt{K_1}\sin(\beta)\tilde{\text{J}}(u) \end{cases} \qquad (31)$$

where $\tilde{\text{J}}(u) = i\text{J}_+(iu),\ \beta = \varphi_1 + \varphi_2 - \varphi_3$.

Since $\text{J}(u)$ is a function with a period of $T$, the only need is to focus on the region: $R = \{(u,\beta)|0 \leq u \leq T/2, -\pi \leq \beta \leq \pi\}$ (named the reduced $(u,\beta)$ plane).

Then let's calculate the stationary points: $(u^{(s)}, \beta^{(s)})$, which satisfying the systems of equations: $\dot{u} = \dot{\beta} = 0$. One can easily get the roots $\beta^{(s)}$ from the first equation of (16):

$$\beta^+ = \pi/2, \beta^- = -\pi/2 \tag{32}$$

Having obtained the value of $\beta^{(s)}$, $u^{(s)}$ can be obtained from solving the first equation of (16), says $u^+ (u^-)$ corresponding to $\beta^+ (\beta^-)$, respectively:

$$\Delta\Gamma - s\sqrt{K_1} \sin(\beta^{(s)})\tilde{J}(u) = 0 \tag{33}$$

For simplicity, set $s=1$. Clearly, to achieve the full frequency conversion from signal to idle wave, "the mixing angle" $u^{(s)}$ needs to change from $u^{(s)} = 0$ to $u^{(s)} = T/2$ during the whole TWM process.

Following the same approach, set the nonlinear adiabaticity condition of OPA in terms of $u^{(s)}$ to be:

$$\left|\frac{du^{(s)}}{d\xi}\right| \ll |\Omega| \tag{34}$$

where $\Omega = \sqrt{-K_1 \tilde{J}'(u^{(s)})}$. For the case when $m=0$, the adiabaticity condition of SFG in terms of $u^{(s)}$ writes:

$$\left|\frac{du^{(s)}}{d\xi}\right| \ll \sqrt{\Delta\Gamma\sqrt{\Delta\Gamma^2 + 4K_1}} \tag{35}$$

Recall that the initial intensity of pump light is much larger that the signal, the adiabaticity condition writes (see Section II.B):

$$|d\theta/dz| \ll \sqrt{\Delta k^2 - q^2}, for\, |\Delta k| > q \tag{36}$$

Set nonlinear adiabaticity parameter to be $r_{nl}$:

$$r_{nl} = \sqrt{K_1} |\Omega|^{-3} \left|\frac{d\Delta\Gamma}{d\xi}\right| \ll 1 \tag{37}$$

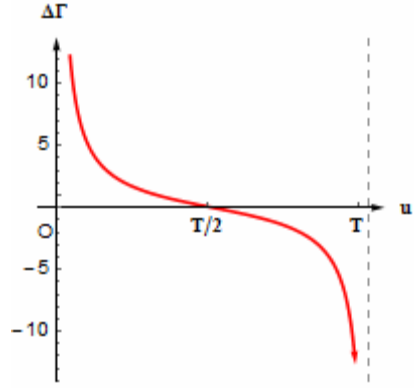

**Figure 7** The adiabatic trajectory $(u^+, \beta^+)$ **of OPA**

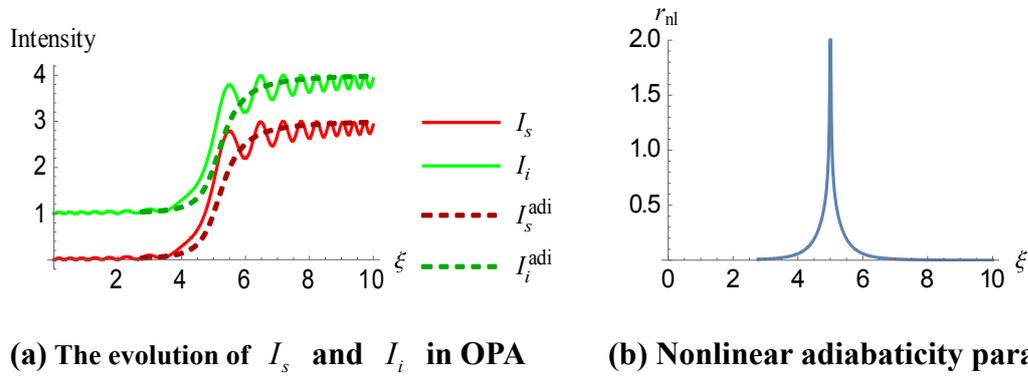

**(a)** The evolution of $I_s$ and $I_i$ in OPA    **(b)** Nonlinear adiabaticity parameter

Figure 8 Adiabatic OPA with $K_1/K_2 = 10$ and $\Delta\Gamma = -4(\xi - 5)$

Finally, according to G. G. Luther et al[10], $[A_1, A_2, A_3]^T$ can be geometrically described by the generalized Bloch vector $\vec{\rho} = \{U, V, W\}$, where

$$U = \text{Re}(A_1 A_2 A_3^*), V = \text{Im}(A_1 A_2 A_3^*), W = \sum_{j=1}^{3} a_j |A_j|^2 \qquad (38)$$

and $a_j$ $(j = 1, 2, 3)$ are any numbers satisfying the relation: $\xi = a_1 + a_2 - a_3 \neq 0$. Therefore, $\vec{\rho}$ falls on the generalized Bloch sphere:

$$U^2 + V^2 = \xi^{-3} \prod_{j=1}^{3}(W_j - W) \qquad (39)$$

where $W_1 = a_2 K_3 - a_3 K_1, W_2 = a_1 K_3 - a_3 K_2, W_3 = a_1 K_1 + a_2 K_2$.

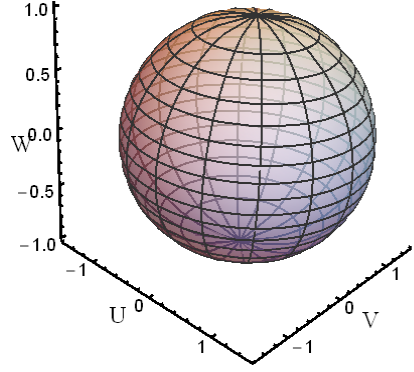

**Figure 9** The generalized Bloch sphere of SFG, with
$$W = I_3 - I_2, I_1^{(0)} = 10 I_2^{(0)}, I_3^{(0)} = 0$$

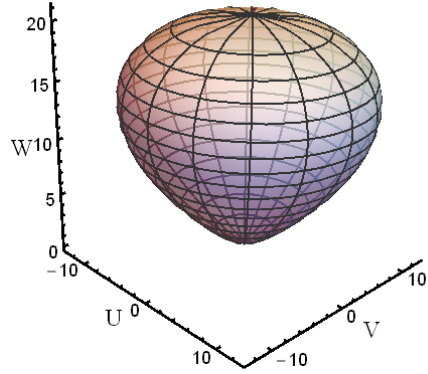

**Figure 10** The generalized Bloch sphere of OPA, with
$$W = I_1 + I_2, I_1^{(0)} = 0, I_3^{(0)} = 10 I_2^{(0)}$$

Obviously, in the undepleted pump approximation, for the case of linear adiabatic SFG(DFG), the normalized amplitude of pump wave is $A_1 \approx A_1^{(0)}$. Set $a_1 = 0, a_2 = -1, a_3 = 1$, generalized Bloch sphere becomes Bloch sphere. And for that of linear adiabatic OPA, the normalized amplitude of pump wave is $A_3 \approx A_3^{(0)}$. Set $a_1 = 1, a_2 = 1, a_3 = 0$, and generalized Bloch sphere becomes pseudo-Bloch sphere.

Moreover, one can represent the generalized Bloch vector $\vec{\rho}$ in terms of $u$ and $\beta$:

$$U = I_1 I_2 I_3 \cos(\beta), V = I_1 I_2 I_3 \sin(\beta), W = \sum_{j=1}^{3} a_j I_j \quad (40)$$

where $I_1$, $I_2$, and $I_3$ are functions of $u$ which have been defined above for different cases. It's easy to see from (17) and (32) that the adiabatic trajectories showed above are just the generatrices of the generalized Bloch sphere, or in other words, the geodesic lines between the bottom and top of the surface.

**IV. Summary**

In conclusion, I have showed that the two cases, i.e. linear adiabatic SFG(DFG) and OPA, are clearly the limiting ones of fully nonlinear TWM process and that the geometric image of the so-called adiabatic basis acts as the shortest paths (geodesic lines) connecting initial state and target state i.e. the generatrices of the generalized Bloch sphere.

The author is glad to receive any useful comments on this article from readers.